# Title:

Enhancing Radiological Diagnosis: A Collaborative Approach Integrating AI and Human Expertise for Visual Miss Correction

# Author list:

**Akash Awasthi**
PhD student
Department of Electrical and Computer Engineering
University of Houston
akashcseklu123@gmail.com

**Ngan Le, Ph.D.**
Assistant Professor
Department of Computer Science & Computer Engineering
University of Arkansas
thile@uark.edu

**Zhigang Deng, Ph.D.**
Moores Professor of Computer Science
Department of Computer Science,
University of Houston, Houston, TX
zdeng4@central.uh.edu

**Carol C. Wu, MD**
Professor
Department of Thoracic Imaging,
Division of Diagnostic Imaging, The University of Texas MD Anderson Cancer Center, Houston
CCWu1@mdanderson.org

**Hien Van Nguyen, Ph.D.**
Associate Professor
Department of Electrical and Computer Engineering
University of Houston
hvnguy35@central.uh.edu

# Corresponding author information:

**Akash Awasthi**
PhD student
Department of Electrical and Computer Engineering
University of Houston

Room no- N368, Cullen College of Engineering Building-1
4222 Martin Luther King Blvd, Houston, TX 77204

akashcseklu123@gmail.com

# Abstract

**Background:** Identifying and correcting perceptual errors in chest radiographs using Human-AI collaboration has not been attempted.

**Purpose:** To develop a collaborative AI system, designed to integrate eye gaze data and radiology reports, improve diagnostic accuracy in chest radiology by identifying perceptual errors and enhancing the decision-making process.

**Materials and Methods:** This retrospective study utilized public datasets REFLACX[27] and EGD-CXR[28] to develop a collaborative AI solution, named Collaborative Radiology Expert (CoRaX). It employs a large multimodal model to analyze image embeddings, eye gaze data, and radiology reports, aiming to rectify perceptual errors in chest radiology. Evaluation focuses on the system's referral-making process, the quality of referrals, and its performance within collaborative diagnostic settings.

**Results:** Proposed system is evaluated on the simulated error dataset comprising 271 samples with 28 % (93 of 332) missed abnormalities corresponding to different regions in the radiographs. Our Proposed system was able to correct 21% (71 of 332) of the errors , 7 %(22 of 312)  still remaining in the dataset. The Referral-Usefulness score metric representing the accuracy of predicted regions  for all True Referrals is 0.63[ 95% CI 0.59, 0.68 ]. The Total-Usefulness score metric, which reflects the diagnostic accuracy of the system's interactions with radiologists, revealed that 84%( 237 of 280) of these interactions (referral and non referral based) have  Total-Usefulness score >0.40.

**Conclusion:** The CoRaX system is designed to collaborate efficiently with radiologists, addressing perceptual errors across various abnormalities. Its potential extends to educational use, offering support in the training of inexperienced radiologists.

# Introduction

AI systems play an increasingly significant role in healthcare decision-making, particularly in diagnostic processes [1,2,3,4,5,6]. However, their integration faces challenges, especially in standalone systems that generally overlook human interaction [ 11,13], potentially leading to decreased diagnostic accuracy due to varying levels of trust among medical professionals[7,8,9]. Collaborative Intelligence offers a promising solution by emphasizing human-AI synergy to enhance accuracy, notably in radiology.

Eye gaze serves as a valuable sensing modality in human-computer interaction [12], fostering dynamic collaboration between radiologists and AI systems [10]. Existing studies have shown a direct correlation between eye movements and radiologist's diagnostic decisions [14,15,16], making eye-tracking data crucial for understanding the visual search process [17,18,19,20,21,22,23] and reducing diagnostic errors[15,16]. Furthermore, eye gaze recording is non-intrusive and can be seamlessly integrated into clinical workflows.

Our study introduces a Collaborative Radiology Expert  (CoRaX), a system utilizing a multimodal model to process image embeddings, eye gaze data, and radiology reports. CoRaX aims to identify and rectify perceptual errors by analyzing radiologists' interactions with radiographs. It consists of two key modules: MAF, focusing on identifying missed abnormalities in radiology reports, and STARE, predicting the spatial and temporal grounding of abnormalities to improve diagnostic accuracy. CoRaX's effectiveness was evaluated in reducing misdiagnosis and improving time efficiency, compared to standalone AI systems and traditional methods.

This study specifically addresses visual misses caused by recognition and decision issues, categorizing them as perceptual errors when radiologists fail to mention or recognize abnormalities[24,25]. Illustrated in Figure 1, CoRaX operates as a post-interpretation system, where radiologists submit X-ray images, reports, and eye gaze data. CoRaX then generates referrals for further assessment by radiologists, forming a collaborative framework between them and the system, with eye gaze data aiding in understanding radiologists' cognitive processes.

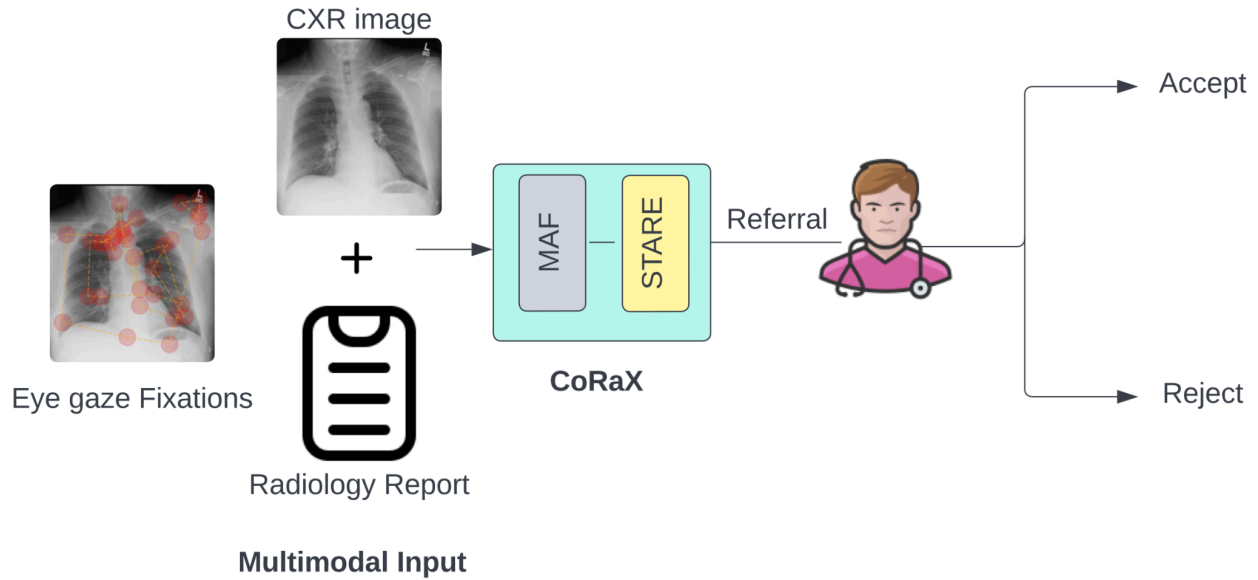

**Figure 1:** An overview of our innovative collaborative system, CoRaX. Our system seamlessly integrates radiology reports, eye gaze data, and chest X-ray (CXR) images to offer targeted recommendations. Then the radiologist uses these recommendations and either accepts them or rejects them.

## Materials and Methods:

**IRB statement:** The study procedures were reviewed and approved by the Institutional Review Board (IRB). This study was conducted in accordance with the ethical standards of the responsible committee on human experimentation and with the Helsinki Declaration of 1975, as revised in 2000. The IRB approval number for this study is STUDY00003659.

**System Overview:** Illustrated in Figure 2, CoRaX comprises two crucial modules: 1) Missed Abnormality Finder (MAF) and 2) Spatio-Temporal Abnormal Region Extractor (STARE). Additionally, it involves specific set operations to identify perceptual errors. Our primary focus is not on generating full radiology reports but on identifying key findings or abnormalities mentioned in the ChexPert dataset. We aim to highlight these significant abnormalities if overlooked by the radiologist.

The MAF module is tasked with summarizing the radiology report and identifying any missing abnormalities. If the radiologist fails to diagnose an abnormality in the CXR image, this module appends the missing abnormality to the summarized radiology report. Refer to the detailed description of the MAF module in the supplementary file.

The STARE module is a pivotal component of our system. It predicts the temporal grounding (i.e., time stamp) for each abnormality in the corrected and summarized radiology report.

Inspired by a dense video captioning task in computer vision [29,30]. STARE takes the MAF module's output (i.e., a corrected and summarized radiology report) and the eye gaze video to predict timestamps for each abnormality in the report, as depicted in Figure 2. The detailed overview of the STARE module is presented in the supplementary file.

To identify missing abnormalities in the actual radiology report, we compare it with the output of the ChexpertLabler [26] module, an NLP tool that summarizes the radiology report into 14 labels [26]. In Figure 2, Set A represents the actual summarized radiology report, and set B represents the output of the STARE module, excluding timesteps for the analysis of the set difference. The set difference reveals missing abnormalities in the radiology report. Set B has a cardinality greater than or equal to that of set A, aligning with our system's focus on correcting perceptual errors or finding missing diagnoses. Our system works well for the 6 abnormalities defined in Table 1.

The set difference result reveals missing abnormalities, and we extract the corresponding timesteps from the STARE module output. Using these timestamps, we identify fixation points between them, representing frames in the video. If this time interval spans multiple frames, we compute the mean of the image frames and merge them into a single image, referred to as the region of interest for the missing abnormality. But we also provide an option in the system if the user wants to have a more detailed look then the user can also pool all the fixation points between the extracted time step into a single heatmap called static heatmap without taking a mean image. This consolidated image and the missed abnormality serve as the referral produced by our system.

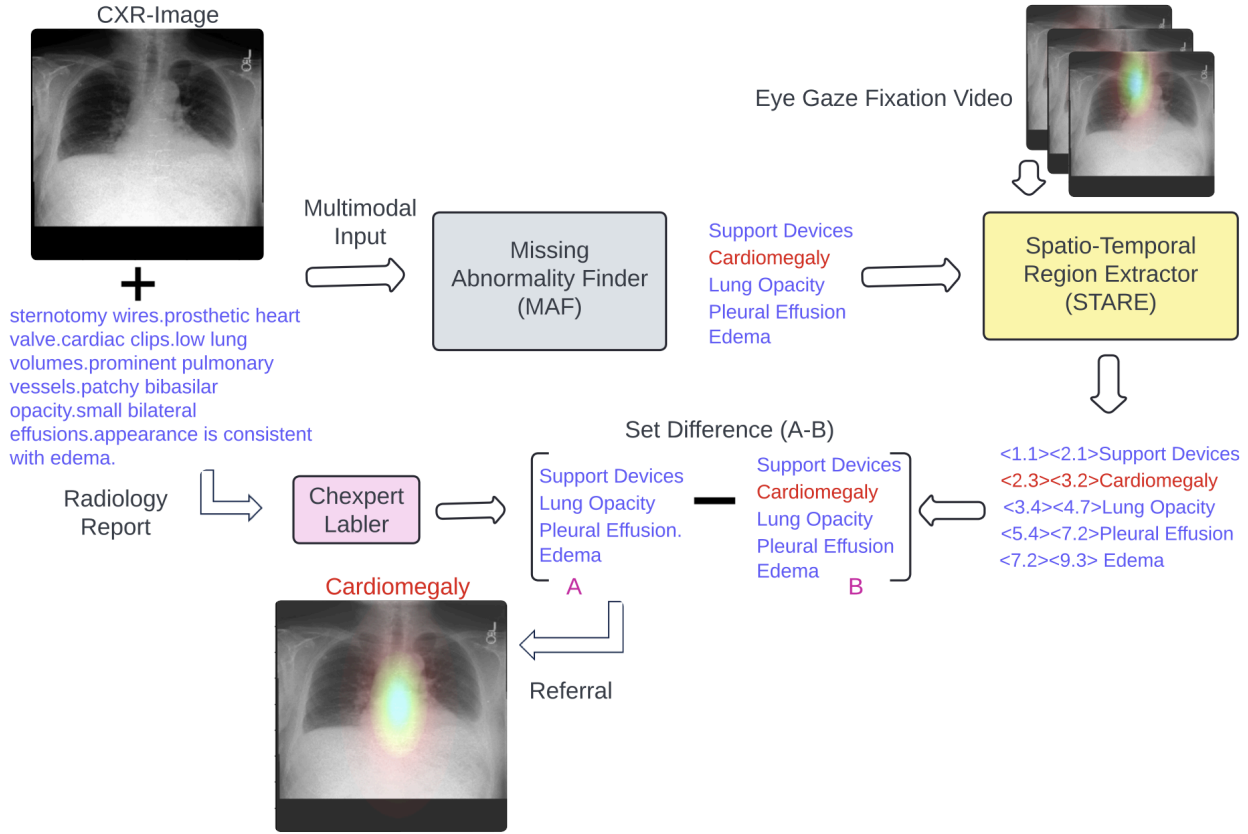

**Figure 2:** Overview of CoRaX architecture, consisting of two main modules. The MAF module is dedicated to identifying abnormalities that may have been missed, while the STARE module focuses on precisely locating the corresponding regions of interest within the diagnostic process.

**Dataset:** For this study, we utilized the EGD-CXR[28] and REFLACX[27] datasets. The EGD-CXR dataset comprises 1,071 chest X-ray images reviewed by radiologists using an eye-tracking system. Meanwhile, the REFLACX dataset encompasses 2440 cases with synchronized eye-tracking and speech transcription pairs, annotated by radiologists. We generated fixation heatmaps overlaid on X-ray images by leveraging eye gaze data, providing a dynamic representation of the gaze movements over the CXR images. The two datasets belong to two different radiologists and we combined them and created a training set and a test set to train the STARE module. The speech transcription data obtained from EGD-CXR and REFLACX contains detailed radiology reports with word alignments for CXR images. By merging the transcriptions from REFLCAX and EGD-CXR, we generated a final JSON file after preprocessing that includes comprehensive reports and associated timesteps. This compilation is crucial for training the STARE module.

**Training and Testing**: To effectively manage computational costs and utilize a significant portion of publicly available datasets, our proposed system employs a strategy of training each module separately. Specifically, ChexFormer undergoes training on the Chexpert dataset[26] and is subsequently tested on its corresponding test set. Meanwhile, the STARE module is trained on a merged dataset comprising the REFLACX and EGD-CXR datasets, resulting in a total of 3511 samples. Among these samples, 2,969 are designated for training, 271 for validation, and 271 for testing purposes.

The training dataset for the STARE module encompasses fixation heatmap videos, summarized radiology reports, and temporally grounded abnormality sequences extracted from the reports, serving as the ground truth. For optimization, we employ the Adam optimizer with a batch size of 2 for both validation and training stages. Fine-tuning of the STARE module requires 300 epochs, with each batch requiring a day of computation on 8 Tesla GPUs.

However, the principal aim of our system is to identify and rectify perceptual errors. To accomplish this goal, we generate an error dataset by simulating the errors from the aforementioned test set. Subsequently, CoRaX is evaluated on this error dataset to assess its effectiveness in identifying and rectifying perceptual errors.

**Error dataset:** The summarized radiology report of the aforementioned test set, which comprises 271 samples, have been altered deliberately to introduce diagnostic errors across different abnormalities in different regions of radiographs. Approximately 28% of the cases within the test set were deliberately adjusted to simulate errors. These alterations were made randomly, without following specific criteria for changing abnormalities as our goal is to ensure that the system benefits both inexperienced radiologists, such as residents in training, and experienced radiologists.

The primary aim of this study is to identify perceptual errors and rectify them. When we mention "altered," it means that if a specific case originally includes a particular abnormality, we either mask it or alter it with negation. Both scenarios fall under the category of perceptual errors resulting from recognition or decision-making. Consequently, specific abnormalities listed in Table 1 have been randomly modified in the test data at varying percentages, as indicated below.

| Abnormality | Percentage ( Error cases/ total cases ) % |
|---|---|
| Cardiomegaly | 15.3(10/65) |
| Pleural Effusion | 23.0(15/65) |
| Atelectasis | 42.0 (23/54) |
| Lung Opacity | 27.0(26/94) |
| Edema | 35.0(19/54) |

**Table 1: This table summarizes the modified cases where errors have been introduced into the test set. The introduced errors correspond to the specified abnormalities, and the table includes the distribution of these introduced errors.**

## Statistical Analysis:

All analyses are performed using the Python version 3.8. Referral-Usefulness score metric which is based on Jaccard Index also called IoU score is used to assess the system's ability for predicting the region of interest for each missed abnormality. 95 % CI is calculated for all true referrals based on normal distribution (sample size >30). 95 % CI is calculated for all abnormality specific true referrals based on t distribution (sample size <30). Cumulative Distribution function is used to analyze the distribution of all interactions (referral and non referral).

# Results

**Illustration of referrals:** We present some case examples alongside the output generated by our system in figure 3. In Figure 3, the layout comprises three columns: the first column displays the actual CXR image, the second column presents the incorrect radiology report derived from simulated error data with its correct counterpart, and the third column exhibits the referrals generated by our system following interpretation. A check mark indicates the radiologist's acceptance of the referral, while a cross mark indicates rejection.

For instance, in Case 2, our system demonstrates its ability to identify multiple missed abnormalities and generate corresponding referrals. It identifies pleural effusion, edema and lung opacity overlooked by the radiologist and highlights two distinct regions indicating the presence of effusion,edema and opacity. This observation suggests the possibility that the radiologist may have missed diagnosing both left and right effusions.

| CXR Image | Radiology Report | Referrals |
|---|---|---|

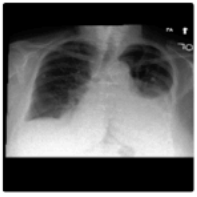

**Incorrect Radiology reprot**
so sternotomy wires. left pleural effusion. patchy right infrahilar opacity.could represent edema or infection.small right effusion.

**Correct Radiology reprot**
so sternotomy wires.moderate cardiomegaly. left pleural effusion with atelectasis and consolidation.patchy right infrahilar opacity.could represent edema or infection.small right effusion.

Cardiomegaly · Atelectasis

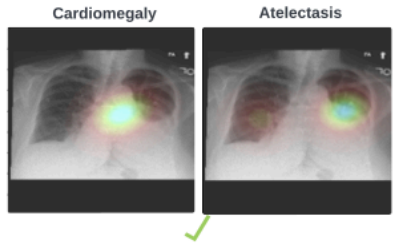

✓

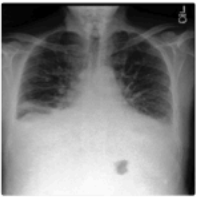

**Incorrect Radiology reprot**
low lung volumes.prominent heart.prominent pulmonary vessels.atelectasis

**Correct Radiology reprot**
low lung volumes.prominent heart.prominent pulmonary vessels.band of atelectasis in the right base.likely small effusions, with patchy opacity, edema and atelectasis are the main considerations.

Edema · Effusion · Lung Opacity

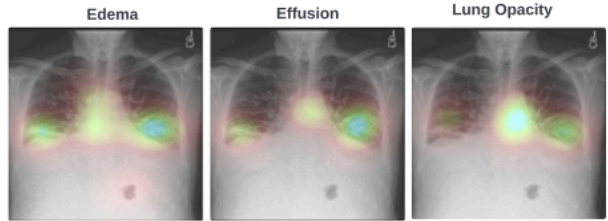

✓ ✓ ✓

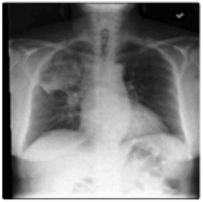

**Incorrect Radiology reprot**
prominent pulmonary vasculature.prominent heart.

**Correct Radiology reprot**
dense opacity in the right mid and upper lung which could represent loculated fluid versus pneumonia.prominent pulmonary vasculature.prominent heart.

Lung Opacity

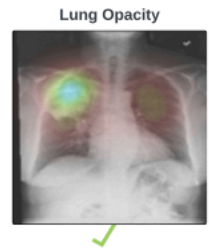

✓

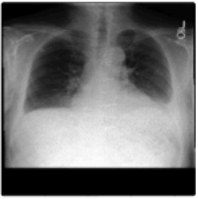

**Incorrect Radiology reprot**
sternotomy wires.prosthetic heart valve.cardiac clips.low lung volumes.prominent pulmonary vessels.patchy bibasilar opacity.small bilateral effusions.appearance is consistent with edema.

**Correct Radiology reprot**
sternotomy wires.prosthetic heart valve.cardiac clips.cardiomegaly.low lung volumes.prominent pulmonary vessels.patchy bibasilar opacity.small bilateral effusions.appearance is consistent with edema.

Cardiomegaly

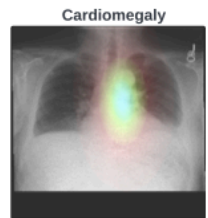

✓

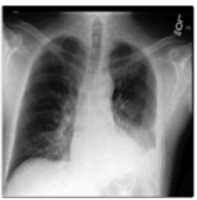

**Incorrect Radiology reprot**
mildly enlarged heart.

**Correct Radiology reprot**
mildly enlarged heart.there is a left effusion moderate in size with some atelectasis or consolidation.

Effusion

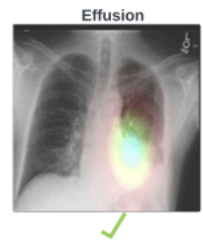

✓

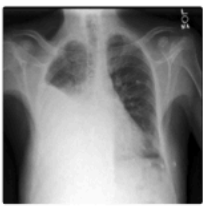

**Incorrect Radiology reprot**
sternotomy wires.large right effusion.left lung is mostly clear.there is a small left pleural effusion.

**Correct Radiology reprot**
sternotomy wires.large right effusion.with atelectasis.left lung is mostly clear.there is a small left pleural effusion.

Atelectasis

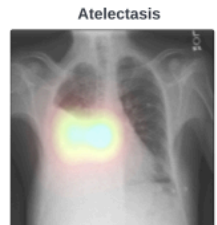

✓

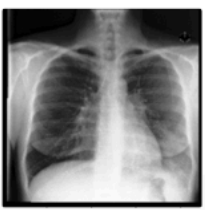

**Incorrect Radiology reprot**
normal heart and mediastinum.lungs are clear.scratch that.right lung is clear.

**Correct Radiology reprot**
normal heart and mediastinum.lungs are clear.scratch that.right lung is clear.

Atelectasis

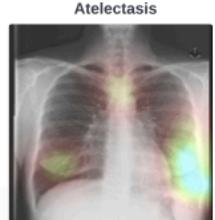

✗

**Figure 3:** Examples of referrals generated by the system for different cases. It includes the actual CXR image, the radiology report (consisting of incorrect radiology report sourced from an error dataset alongside its correct counterpart), and the corresponding referrals made by CoRaX, along with their acceptance or rejection indicated by the blue tick or cross mark respectively. Missing abnormalities are highlighted with blue in the correct radiology report.

**Referral Evaluation:** In this section, we aim to evaluate the usefulness of referrals by identifying missed abnormalities and their corresponding regions of interest.

The evaluation of referrals in identifying missed abnormalities involves assessing how many times such abnormalities are correctly identified and corrected. This assessment includes calculating True Referral (TR), False Referral (FR), False Deferral (FD), and True Deferral (TD) for each abnormality.

**True Referral (TR)** represents the number of cases where missed abnormalities are correctly identified by the CoRaX. It also represents the number of accepted referrals by the radiologists.

**False Referral (FR)** represents the number of cases where the CoRaX over-diagnoses abnormality. It also represents the number of rejected referrals by the radiologists

**False Deferral (FD)** FD represents the number of cases where the model fails to identify missing abnormalities overlooked by radiologists, resulting in incorrect decisions of not making referrals.

**True Deferral ( TD)** TD represents the number of cases where the system correctly decides not to make referrals for abnormalities.

Using TR, FR, FD, and TD, we subsequently compute the Perceptual Error Correction Rate (PECR) and Overdiagnosis Error Rate (ODER), providing a singular metric for understanding the model's performance in identifying errors.

**Perceptual Error Correction Rate (PECR):** The metric measures the system's effectiveness in rectifying perceptual errors for each abnormality, with individual PECR values outlined in Table 1. This metric, resembling the system's recall, is defined as follows:

$$PECR(\%) = \frac{(TR)}{TR+FD} \quad (3)$$

**Over-diagnosis Error Rate (ODER):** This metric measures how frequently the model over-diagnoses abnormalities, indicating the proposed system's error rate:

$$ODER(\%) = \frac{(FR)}{FR+TD} \quad (4)$$

In this simulated dataset, containing around 271 samples, approximately 93 abnormalities are missing. This number 93 represents the number of missed abnormalities (TF +FD) belonging to the different regions in the radiographs , not the number of cases. Because one case can have multiple abnormal conditions missed. As demonstrated in table 2, Cardiomegaly has a PECR of 100%(10 of 10) with ODER of 0.7%(2 of 261), Edema has a PECR of 74%(14 of 19) with ODER of 0.8%(2 of 254), Atelectasis has a PECR of 61%(14 of 23) with ODER of 0.8%(2 of 250),Effusion has a PECR of 67%(10 of 15) with ODER of 0.8%(2 of 258),Lung opacity has a PECR of 88.4%(23 of 26) with ODER of 0.4%(1 of 260).

**Table 2: Breakdowns of CoRaX's performance in identifying visual misses for different**

| Abnormality | True Referrals (TR) | False Deferral (FD) | Perceptual Error Correction Rate (PECR) | False Referral (FR) | True Deferral (TD) | Over-diagnosis Error Rate (ODER) (%) |
|---|---|---|---|---|---|---|
| Cardiomegaly | 10 | 0 | 100.0 | 2 | 259 | 0.7 |
| Edema | 14 | 5 | 74.0 | 2 | 252 | 0.8 |
| Atelectasis | 14 | 9 | 61.0 | 2 | 248 | 0.8 |
| Effusion | 10 | 5 | 67.0 | 2 | 256 | 0.8 |
| Lung Opacity | 23 | 3 | 88.4 | 1 | 259 | 0.4 |

**types of abnormalities. Perceptual error correction rate and over-diagnosis rate are included for complete assessment of CoRax's performance.**

**Evaluation of Referrals based on the region of interest:** The system's ability to identify missing abnormalities and highlight their regions of interest is assessed by measuring the overlap between the true and predicted regions, utilizing the Intersection over Union (IoU) metric[31].

Higher IoU scores indicates better usefulness, indicating reduced confusion for radiologists and offering insights into potential time savings in clinical practice. By effectively directing radiologists to the correct regions of interest for missed abnormalities, the system minimizes the need for them to initiate their search anew.

$$Referral - Usefulness\ score\ =\ I(Referral\ is\ Accepted)\ *\ IoU \quad (5)$$

Where $I$ is an indicator function whose value is 1 if the referral is accepted.

In Figure 4, the detailed evaluation of CoRaX's referral-based interaction is presented across three subfigures. The first two subfigures illustrate Referral-Usefulness score for True Referrals ( TF), indicating the effectiveness of highlighted regions for correctly identified missed abnormalities. The third subfigure depicts the overall performance of the system's referral-based interaction, representing the overall referral usefulness.

In subfigure 4 (a), the distribution of Referral-Usefulness score for all True Referrals is displayed. The mean Referral-Usefulness score for True Referrals (TF) is 0.63[95% CI: 0.59, 0.68].

Subfigure 4(b) provides abnormality-specific Referral-Usefulness scores for true referrals,represented in the strip plots. Notably, Cardiomegaly exhibit mean Referral-Usefulness score of 0.83[95% CI: 0.59, 0.68],Pleural Effusion 0.50[95% CI: 0.35, 0.65] , Lung opacity 0.61[95% CI: 0.54, 0.69] , Edema 0.64[95% CI: 0.53, 0.76] , Atelectasis 0.60[95% CI: 0.53, 0.68 ]

To understand the system's overall performance in referral-based interaction, a Cumulative Distribution Function (CDF) plot of Referral-Usefulness scores for all referrals is presented in Figure 4c. This plot offers insight into the distribution, with  80%(64 of 80) of referrals having a Referral-Usefulness score exceeding 0.2.  However, a sudden drop in the curve indicates 0 usefulness of referrals, with approximately 19%(15 of 80)  being deemed not useful.

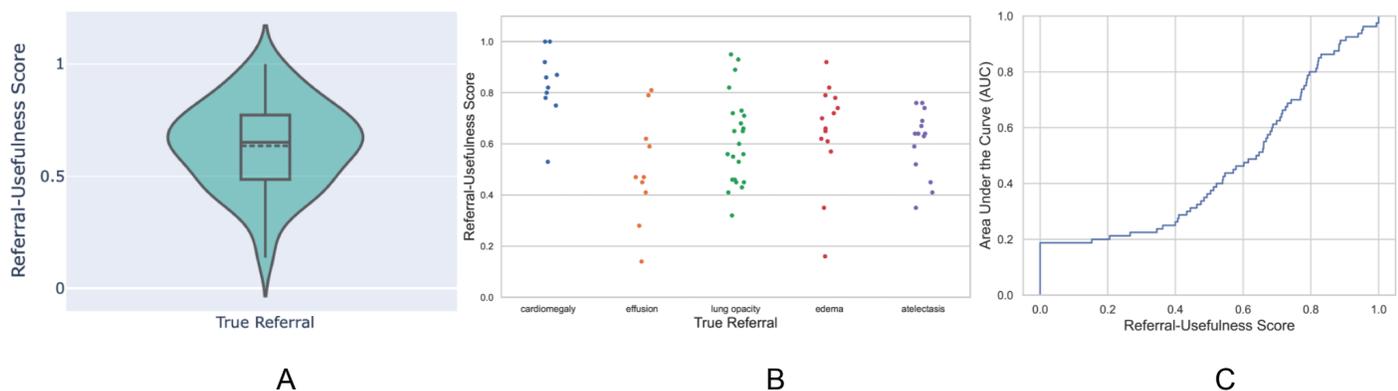

A  B  C

**Figure 4: Evaluation of the proposed system's Referral-based Interaction. The figure comprises three subplots (A, B, C). Plot A illustrates the distribution of Referral-Usefulness scores for all True Referrals (TR), providing insight into their usefulness. Plot B displays the Referral-Usefulness scores for True Referrals (TR) corresponding to each abnormality, highlighting their individual effectiveness. Plot C depicts the overall usefulness of referrals ( TR+FR) based on Referral-Usefulness scores in the Referral-based Interaction.**

**Evaluation of overall interaction and Diagnostic accuracy:** We introduce the Total-Usefulness score metric, which serves as an indicator of the diagnostic accuracy of each interaction between the system and the radiologist. A Total-Usefulness score of 1 indicates a fully beneficial interaction, representing 100% diagnostic accuracy.

Interactions between the system and the radiologist can be classified into two categories: referral-based and non-referral-based. We consider the interaction as non-referral or Deferral based if there is no any referral for a particular case. For non-referral interactions, a score of 1 is assigned when the system refrains from making any referral for a particular case and its decision aligns with the correct diagnosis, ensuring no perceptual errors are overlooked. Conversely, a score of 0 is assigned when the system makes an incorrect deferral, indicating 0% diagnostic accuracy and minimal assistance to the radiologist. In referral-based interactions, we utilize the IoU score to quantify diagnostic accuracy and use them as the Total-Usefulness score.

We present the overall performance of the system in a collaborative environment in Figure 5, representing both referral and deferral-based interactions. It consists of two subfigures. In Figure 5a, approximately 71%(200 of 280) of the interaction result in no referrals, with around 63%(178 of 280) of these decisions being accurate, and 8%(22 of 280) of missed correcting radiologist's perceptual errors. Regarding referral-based interactions, 29%(80 of 280) of interactions are referral based , with around 25%(71 of 280) are correct and 3%(9 of 280) are incorrect, leading to direct rejection by radiologists.

In Figure 5B, we observe that in approximately 84%( 237 of 280) of the interactions( referral and non referral) , the system achieves an Total-Usefulness score exceeding 0.4, indicating significant aid and improvement in diagnostic performance in 84%( 237 of 280) of interactions, benefiting both radiologists and the diagnostic process.

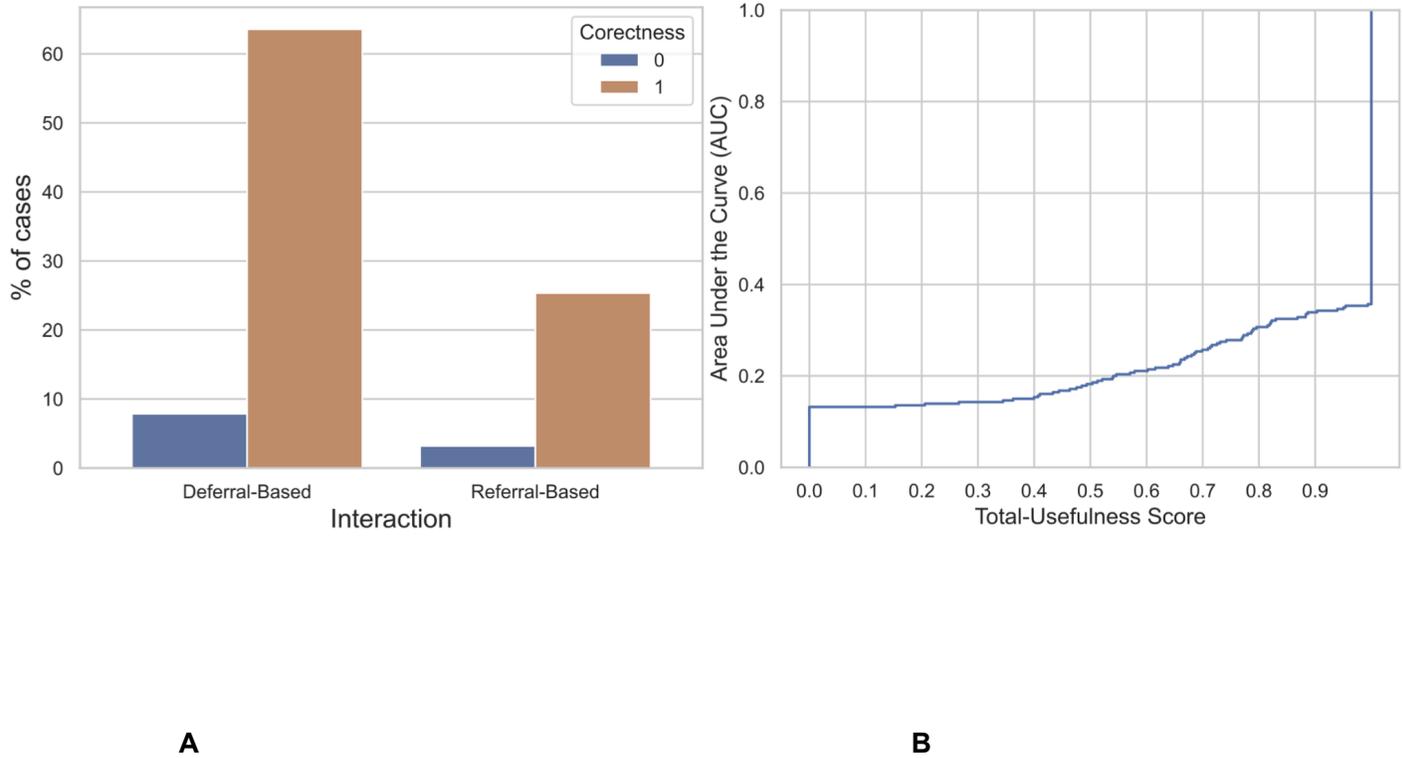

A  B

**Figure 5: Correctness and utility of AI-radiologist interactions. Plot A offers insights into the system's referral frequency, accompanied by the correctness of its referral decisions. Plot B presents the cumulative distribution function of the Total-Usefulness score, reflecting the system's comprehensive performance considering the collaboration between the radiologist and CoRaX.**

# Discussion

Chest radiographs are widely performed globally, but their interpretation is prone to errors, with an estimated 4% incidence of mistakes by radiologists in routine cases [33,34]. The primary error types include perceptual and cognitive errors[24], with perceptual errors constituting the majority (60% to 80%) and often resulting from oversights during the initial interpretation [35]. In response to this we develop a collaborative AI solution, named Collaborative Radiology Expert (CoRaX), that leverages a large multimodal model to process image embeddings, eye gaze data, and radiology reports. The collaborative strategy systematically addresses visual misses, targeting recognition issues and minimizing perceptual errors and makes this system also useful for the training module for the inexperienced radiologists or residents in hospitals. This unique approach, different from standalone AI systems, fosters collaboration with radiologists, marking a notable advancement in diagnosis.

To assess the capability of CoRaX in correcting perceptual error, we created error dataset containing 271 samples with 28%(93/332) of the perceptual error across different abnormalities like Cardiomegaly 15.3 %(10/65), Pleural Effusion 23.0 %(15/65), Atelectasis 42.0 %(23/54),Lung Opacity 27.0% (26/94),Edema 35.0 %(19/54). Our proposed system works as a second reader and makes referrals consisting of missed abnormality and corresponding Region of interest (ROI). Our results demonstrated that CoRaX can correct 100%(10 of 10) of perceptual error for Cardiomegaly, 74%(14 of 19) for Edema, 61%(14 of 23) for Atelectasis ,67%(10 of 15) for Effusion, 88.4%(23 of 26) for Lung opacity. Notably, the proposed system exhibits a notably high Perceptual Error correction rate for cardiomegaly compared to other abnormalities. The Perceptual Error correction rate is largely influenced by the multilabel classifier in our proposed system, and we provide additional details on the classification performance of our classifier in the supplementary section.

To evaluate the usefulness of referrals , referral-usefulness score is calculated which represents the accuracy of predicted regions. Our system demonstrates 0.63[95% CI: 0.59, 0.68] of Referral-Usefulness score for all True Referrals (TF). Cardiomegaly exhibit mean Referral-Usefulness score of 0.83[95% CI: 0.59, 0.68],Pleural Effusion 0.50[95% CI: 0.35, 0.65] , Lung opacity 0.61[95% CI: 0.54, 0.69] , Edema 0.64[95% CI: 0.53, 0.76] , Atelectasis 0.60[95% CI: 0.53, 0.68 ]. To understand the system's overall performance in referral-based interaction, 80%(64 of 80) of referrals having a Referral-Usefulness score exceeding 0.2, suggesting usefulness in most cases. Approximately 19%(15 of 80) of referrals have 0 Referral-Usefulness score representing rejected referrals (FR). Some true referrals may also have a 0 Referral-Usefulness score, indicating 0 usefulness and potential confusion for the radiologist. The system's objective is not solely to identify missed abnormality but also to direct radiologists to the correct region without ambiguity.

Beyond evaluating referral usefulness, it's crucial to assess the system's overall diagnostic precision and intervention efficacy. Introducing the Total-Usefulness score metric, we found that in 84%( 237 of 280) of interactions (referral and non-referral based), CoRaX achieved a Total-Usefulness score exceeding 0.4, significantly aiding and improving diagnostic performance in the majority of interactions (referral and non-referral based). This highlights CoRaX's effectiveness in benefiting both radiologists and the diagnostic process as a whole.

Our study had limitations. Firstly, in the simulated error dataset, we deliberately introduced perceptual errors. Despite efforts to ensure these errors mirrored common missed patterns through manual examination of erroneous samples, we were unable to fully encompass the wide spectrum of perceptual errors present in genuine radiology reports[25]. Secondly, there was a minor misalignment between eye gaze movements and report transcription, resulting in slight inherent errors in the regions predicted by the STARE module.

 In Conclusion, the development of CoRaX represents a significant advance in the integration of AI in medical diagnostics, particularly in radiology. By leveraging the collaborative intelligence model, CoRaX systematically addresses the limitations of standalone AI systems, demonstrating a notable improvement in diagnostic accuracy and efficiency. This study underscores the potential of human-AI collaboration in enhancing healthcare outcomes, suggesting a promising direction for future research and implementation in clinical settings.